# Globally disruptive events show predictable timing patterns

Michael P. Gillman

Evolution and Ecology Research Group

School of Life Sciences, University of Lincoln

Brayford Pool, Lincoln, LN6 7TS, United Kingdom

Corresponding author, mgillman@lincoln.ac.uk

Hilary E. Erenler

Landscape and Biodiversity Research Group

School of Science and Technology, University of Northampton

Newton Building, Northampton, NN2 6JD, United Kingdom

**Abstract**

Globally disruptive events include asteroid/comet impacts, large igneous provinces and glaciations, all of which have been considered as contributors to mass extinctions. Understanding the overall relationship between the timings of the largest extinctions and their potential proximal causes remains one of science's great unsolved mysteries. Cycles of about 60 million years in both fossil diversity and environmental data suggest external drivers such as the passage of the Solar System through the galactic plane. While cyclic phenomena are recognised statistically, a lack of coherent mechanisms and a failure to link key events has hampered wider acceptance of multi-million year periodicity and its relevance to earth science and evolution. The generation of a robust predictive model of timings, with a clear plausible primary mechanism, would signal a paradigm shift. Here, we present a model of the timings of globally disruptive events and a possible explanation of their ultimate cause. The proposed model is a symmetrical pattern of 63 million-year sequences around a central value, interpreted as the occurrence of events along, and parallel to, the galactic midplane. The symmetry is consistent with multiple dark matter disks, aligned parallel to the midplane. One implication of the precise pattern of timings and the underlying physical model is the ability to predict future events, such as a major extinction in one to two million years.

**Introduction**

Cycles on time-scales of tens of millions of years have been observed across multiple datasets relating to the fossil record, oxygen and strontium isotopes, large igneous provinces (LIPs) and impact craters (Raup and Sepkoski 1984, Rohde and Muller 2005, Melott *et al.* 2012, Rampino and Prokoph 2013, Melott and Bambach 2014, Shaviv *et al.* 2014, Rampino 2015, Rampino and Caldeira 2015). A highly statistically-significant cycle of about 62 Myr detected in the fossil data has been found in LIP volumes and strontium isotopes, along with cycles of approximately 27 and 140-175 Myr (Rohde and Muller 2005, Melott *et al.* 2012, Prokoph *et al.* 2013, Melott and Bambach 2014,

Rampino and Caldeira 2015). Combined studies of fossil records with possible causes/proxies suggest similarities in phase (Melott *et al.* 2012, Prokoph *et al.* 2013). Other studies have revealed close temporal links between extinctions and their potential triggering events, most notably the coincidence of the Chicxulub impact with the end-Cretaceous extinction (Renne *et al.* 2013), marking the final demise of the dinosaurs.

At present there is no clear explanation, or even consensus, regarding multi-million year cycles on the basis of internal Earth processes (Melott *et al.* 2012, but see comments in Courtillot and Olson 2007). More promising, however, are explanations which are based on circumstances relating to the passage of the Solar System through the Milky Way; for instance during galactic plane oscillation and/or galactic epicycles (see e.g. Rampino and Stothers 1984, Gies and Helsel 2005, Rohde and Muller 2005, Medvedev and Melott 2007, Melott 2008, Randall and Reece 2014, Shaviv *et al.* 2014). The environmental changes which the Solar System undergoes during its journey around the galaxy have, in turn, been linked to gamma-ray bursts, variations in the flux of cosmic rays and the distribution of dark matter (Medvedev and Melott 2007, Thomas 2009, Randall and Reece 2014, Shaviv *et al*. 2014, Rampino 2015).

Here we investigate the pattern of timings of key geological and biological events on Earth. These encompass the most severe extinctions, the most extensive LIPs, the largest asteroid/comet impacts and the Neoproterozoic glaciations. All of these events were either global in extent or affected widely-dispersed ecosystems. Our aim is to capture the largest well-dated events across a range of categories relevant to life on Earth in a single analysis. We consider whether the events are uniformly distributed in time and, if not, the form of the resultant pattern and how it may be explained by physical processes.

**Methods**

The data consist of the ages of 24 impacts with craters of diameter greater than 20km, the 15 largest LIPs with minimum area of 1 million km$^2$, the eight most severe extinctions and the three Neoproterozoic glaciations (see Table 1 and Supplementary Information for more details on these events). The minimum temporal resolution is considered by combining events whose ages overlap within errors, thereby avoiding pseudoreplication, e.g. two impacts with a common origin being counted separately, or a glaciation resulting from a LIP being counted as two events. This gives a total of 38 discrete events or event combinations, approximately 90% of which have errors of less than 2.5 Myr.

The initial hypothesis to be tested was whether the event data were non-uniform on a specific period(s). Data were analysed using circular statistics in R 3.2.1. (R Core Team 2015) to test for departures from a uniform distribution. Specifically, we employed Kuiper's one-sample test of uniformity in the CircStats package (Lund and Agostinelli 2012). This was checked against the Rayleigh test of uniformity which has an alternative hypothesis of a unimodal distribution and gave very similar results. Kuiper's test is a non-parametric test close to the Kolmogorov-Smirnov test, with a test statistic based on the magnitudes of the largest differences between the observed and uniform null distribution. Kuiper's test is responsive to changes in the tails of the distribution and the median values, making it invariant under cyclic transformations. Analyses were performed at different integer modulo values from 20-80 Myr inclusive, to cover the full range of possible <100 Myr cycles discussed in the literature, and then fine-tuned to 0.01 Myr values around the significant cycle length. Significant departures from uniformity imply clusters or aggregations of events which then allows the location (through the significant cycle time), and the event composition of the clusters, to be determined.

Two approaches were used in the detection of clusters. The first used a moving time-window of up to five million years. Potential clusters were tested against a null hypothesis of independent occurrence under a Poisson distribution. The second method involved analysis of the distribution of events within and outside the time window defined by the Cretaceous long normal superchron (see Results for comparison with Kiaman superchron timings). Superchrons, which are extended periods of non-reversal of the Earth's polarity, were chosen as there has been discussion of the relationship of their timings to mass extinctions (Courtillot and Olson 2007, Wendler 2004) and they therefore represented both an *a priori* selection of timings, in contrast to the *post hoc* identification of clusters of events via the Poisson analysis, and a different scale of potential clustering (30-40 Myr rather than <5 Myr). In this case the observed distribution of events was contrasted with a null hypothesis of the numbers being proportional to the time in each window (using a log-likelihood ratio test, G-test, with the outcome assessed against a chi-squared distribution, Sokal and Rohlf 1981). The initial detection of periodicity and subsequent detection of clusters using Poisson analysis was independent of superchron ages or durations.

**Results**

*Non-uniform distribution of events*

The circular statistical analyses revealed significant ($P<0.01$) deviations of the event ages from the null hypothesis of uniform distribution, within a narrow range from 62.9 to 63.6 Myr inclusive. The periods are consistent with the strong 62 Myr signal from the fossil record, LIP volumes and strontium isotopes. The lowest probabilities (highest test-statistic values) occurred at 63.27 Myr (Figure 1) which we use as the cycle period for modelling and to determine equivalent ages (EA, Table 1), i.e., event ages that occupy the same position in the 63.27 Myr cycle (Figure 2).

Four significant clusters of events were detected across approximately one quarter of the 63.27 Myr cycle (see Figure 2 and Table 2 in Supplementary Information). The four clusters fall outside of the equivalent ages of the Cretaceous long normal and Kiaman superchrons. The overall density of events in the region outside the Cretaceous superchron (which lies within the Kiaman start and end equivalent ages) was significantly higher than the density within the superchron region (26 events outside, 12 events within, G=12.0, 1 d.f., P=0.0005).

*Symmetrical model of timings*

We propose a symmetrical model of timings (Table 2) based on the distribution of clusters, their relationship to stratigraphic boundaries (Cohen *et al.* 2013) and the linkage of similar events (see below). The central sequence of ages is set at 66.0 Ma ± $n$ x 63.27 Myr (where $n$ is an integer) with two pairs of sequences either side at 4.5 Myr and 10.2 Myr from the centre, parameterized by stratigraphic boundaries (Table 2, Figure 2). The 10.2 Myr clusters of events are observed to occur within a range of 8.9 to 11.1 Myr from the central value. The differences from the central value, such as 4.5 Myr, are referred to as equivalent differences or ED (Tables 1 and 2).

The central sequence links events considered in the present analysis, such as the end-Cretaceous (66.04 ± 0.043 Ma), the end-Ordovician (445 ± <2 Ma) and the Marinoan glaciation (635.6 ± 0.5 Ma), with other globally important events such as the expansion of ice/colder climate about 3 Ma in the southern hemisphere (Petrick *et al.* 2015), the Hauterivian-Barremian boundary (~129.4 Ma, coinciding with a rapid sea level change, Lukeneder 2012) and the middle-upper Devonian transition (382.7 ± 1.6 Ma, a major extinction event linked to sea level rise, McGhee *et al.* 2013).

The first pair of sequences, with timing occurring 4.5 Myr away from the central value, includes three major extinctions (the end-Permian at 251.9 ± <0.2 Ma, end-Serpukhovian at 323.2 ± 0.4 Ma and end-Capitanian at 260.3 ± <1.0 Ma), together with the largest comet impact at 2023 ± 4 Ma (see clusters at ± 4.5 Myr, Figure 2). The second pair of sequences occurring 10.2 Myr from the central value agrees with the observed ages of the Rochechouart and Lappajärvi impacts and the Gaskiers glaciation (76.2 ± 0.29, 202.7 ± 2.2 and 582.4 ± 0.5 Ma respectively, Figure 2, Table 2). This outer pair of sequences brings together events characterised by a variety of globally important and potentially causally related phenomena, including ocean anoxia, rapid global climate shifts and pronounced isotope excursions, which in turn impact on/reflect marked changes in biodiversity, e.g. during the Palaeocene-Eocene thermal maximum (~56 Ma, Yamaguchi and Norris 2015), the Toarcian oceanic anoxic event (183.22 ± 0.26 to 181.99 ± 0.13 Ma, Sell *et al.* 2014), the end Frasnian extinction (stratigraphic boundary at 372.2 ± 1.6 Ma and extinction possibly at 373.9 Ma, De Vleeschouwer and Parnell 2014), the lower-middle Devonian boundary (393.3 ± 1.2 Ma, Elrick *et al.* 2009) and the late Cambrian SPICE event (~499 Ma, Dahl *et al.* 2014).

The range of ED from 11.09 to 8.9 Myr for the outer pair of sequences (Table 2) is consistent with the observed duration of events in well-studied strata, e.g., the predicted ages of 203.6 to 201.4 Ma agree with end-Triassic magmatism from 203.71 ± 0.11 to 201.32 ± 0.13 Ma (although also earlier at ~205.4 Ma, Wotzlaw *et al.* 2014), culminating in the timing of the end-Triassic extinction at 201.4 ± 0.2 Ma (Ikeda and Tada 2014; Wotzlaw *et al.* 2014) and the major expansion of ice at approximately 13.9 Ma followed by declines in $\delta^{13}C$ between 13.5 and 12.7 Ma (Holbourn *et al.* 2013).

The sequences are tentatively extended back to the earliest solar events of calcium-aluminium-rich inclusions (CAIs) formation at 4567.3 ± 0.16 Ma (Connelly *et al*. 2012), equivalent to 11.86 Ma, and chondrules from the Gujba chondrite with an age of 4562.49 ± 0.21 Ma (Bollard *et al*. 2015) equivalent to 7.05 Ma, i.e., within 0.2 Myr of the Tortonian-Messinian boundary (Table 2).

While we focus on symmetrical patterns of ~63 Myr cycles, we note that the proposed model may also help to explain previously documented 27 Myr cycles. For example, Rampino and Caldeira (2015) have shown a significant 27.0 Myr cycle in extinctions with a most recent maximum of 11.8 Ma. The sequence of 27 Myr peaks over the last 260 Myr is predicted at 11.8, 38.8, 65.8, 92.8, 119.8,

146.8, 173.8, 200.8, 227.8 and 254.8 Ma. This is consistent with the values of 11.63 (ED of +8.9), 66.0 (0), 120.4 (-8.9), 201.4 (+8.9) and 255.8 (0) in Table 2.

*Formulation of a physical model*

Generation of a physical model to explain this pattern is both facilitated and constrained by five results and observations relating to the present analysis. First and foremost, the most recent central value in our model of timings, namely 2.73 Ma (66.0 Ma minus 63.27 Myr), is very close to the estimated timing of the last galactic midplane crossing by the Solar System at 2.8 Ma (Joshi 2007, Shaviv *et al.* 2014). Second, if galactic midplane crossings occur every 63 Myr, the symmetrical model is consistent with equal timings before and after galactic midplane crossings. When embedded into a physical model featuring oscillations around the galactic midplane, the simplest interpretation is that 63 Myr corresponds to half a period of oscillation. Third, the clusters include events with observed ages up to 2 Ga. Assuming the model can reliably be extrapolated that far, the timings may also be consistent with the earliest observed Solar System calcium-aluminium-rich inclusions. Fourth, the inferred clusters of events often relate to relatively rapid changes on timescales of less than one million years against backdrops of low change, e.g., episodes of environmental change through the early Cretaceous (Follmi 2012). Finally, periodic vertical oscillations of the Sun around the galactic plane are assumed to be a reasonable approximation to its true motion, despite the potential for perturbation by a series of agents.

It has been shown elsewhere that the presence of a thin disk of dark matter can address shortcomings faced by galactic midplane crossing models based on baryonic matter (Randall and Reece 2014). If dark matter is an important driver of perturbations, it follows that the symmetrical pattern of timings could be produced by multiple dark disks parallel to the galactic midplane, which may occur as an extension of the models of Fan *et al.* (2013a,b) and Foot and Vagnozzi (2015a,b). It is envisaged that the small fraction (10% or less) of dark matter being dissipative resides in separate unbroken Abelian hidden sectors. In its simplest form, each of these hidden sectors features a massless dark photon (responsible for dissipative interactions which cool the dark matter to a disk), a light coolant fermion and a second heavy fermion. These models are consistent with bounds from the Cosmic Microwave Background at early times, and the Large-Scale Structure at late times. Furthermore, current limits on the dark matter self-interaction cross-section from galaxy cluster collisions (Harvey *et al.* 2015) do not overly restrict the available parameter space for such models. It is expected that the thin disks would be approximately parallel, aligned by gravitational interactions and with similar initial angular momenta. (S. Vagnozzi pers. comm.). As the Sun oscillates around the galactic midplane, the vertical density gradient caused by the varying gravitational attraction of the underlying disks of different scale heights is responsible for tidal forces. These in turn may perturb the Oort Cloud and the asteroid belt, disrupt Milankovitch cycles and affect the Earth tide (increasing/prolonging LIP activity), all of which have the potential to contribute to rapid changes in global climate.

**Discussion**

The model of timings proposed here is novel in two important respects. First, it explicitly combines different geobiological phenomena into a single analysis. Second, it reveals a repeating symmetrical pattern of events rather than a single cyclical process. The model has major implications for unravelling extinction mechanisms, exploring the history of the Earth, predicting future events and determining the structure of dark and luminous matter around the galactic plane.

Understanding the causes of extinction is facilitated by linking temporally disparate, and apparently unrelated, events via the identified ~63 Myr sequences and considering them in the context of an underlying physical model. For example, the role of rapid global glaciations (and, more generally, rapid global climate change) in mass extinctions is highlighted by recognising that the Marinoan glacigenic formations and start of the Hirnantian (with coincident glaciation and extinction, Melchin *et al.* 2013) have equivalent ages within one million years of the end-Cretaceous (Table 1). The latter, best known for the coincidence of the Chicxulub impact and the Deccan Traps, was also a period which culminated in global temperature drops (Renne *et al.* 2013).

If the sequences of past events considered here correspond to the interception of disks of dark matter by the travelling Solar System, then the chronology of its constituents represent a powerful tool to calibrate and interpret the physical structure of the galaxy. Conversely, the increasingly detailed and accurately dated stratigraphic record of environmental change on Earth provides a variety of opportunities to test hypotheses of related events in the 63 Myr sequences. Intriguingly, the chronology may extend back to the early evolution of the Solar System itself, with the gravitational collapse of dense clumps of molecular cloud and formation of a proto-sun occurring approximately 1 Myr before CAIs (Amelin and Ireland 2013), thereby locating the proto-sun age in the significant cluster of 10.2 Myr (equivalent to 4568.37 Ma) before the central value. Furthermore, the dark matter model has the potential to undergo close scrutiny in the coming years, through precise measurements of stellar kinematics by the Gaia satellite (ESA 2015, Randall and Reece 2014).

The impetus for further work is not mere curiosity. The repeating pattern of mass extinctions and associated events are expected to continue into the future, with the next cluster in approximately 1.8 million years. A fuller understanding of the patterns, durations and causes of catastrophic events will determine whether that seemingly distant cluster is treated as an irrelevant hypothetical possibility or a phenomenon worthy of consideration by future human generations.

**Acknowledgements**

Thanks to Sunny Vagnozzi for extensive discussion on the manuscript and sharing his ideas on the underlying physical model. Thanks also to Adrian Melott, Simon Kelley and Stuart Humphries for helpful comments on earlier versions of the manuscript and Ed Gillman for discussions through several iterations of the manuscript. The manuscript benefitted greatly from the comments of two anonymous referees.

| Event type | Event name(s) | Absolute age (Ma) | EA (Ma) | ED (Myr) |
|---|---|---|---|---|
| I | Kara-Kul | 2.5 | 65.77 | -0.23 |
| I/[GC] | Ries/Monterey | 14.2 | 77.47 | +11.47 |
| LIP | Afar | 31 | 94.27 | +28.27 |
| I | Chesapeake, Popigai | 35.7 | 35.7 | -30.3 |
| I | Kamensk, Montagnais | 50.4 | 50.4 | -15.6 |
| LIP | NAVP | 61.5 | 61.5 | -4.5 |
| I/E/LIP | Chicxulub, Boltysh/end-Cretaceous/Deccan Traps | 65.9 | 65.9 | -0.1 |
| I | Kara | 70.3 | 70.3 | +4.3 |
| I | Lappajärvi | 76.2 | 76.2 | +10.2 |
| LIP | Madagascar | 92 | 92 | +26.0 |
| I/LIP | Tookoonooka/Ontong Java | 125 | 61.73 | -4.27 |
| LIP | Parana-Etendeka | 134.3 | 71.03 | +5.03 |
| I | Mjolnir/Gosses Bluff | 142.3 | 79.03 | +13.03 |
| I | Morokweng | 145.2 | 81.93 | +15.93 |
| I | Puchezh-Katunki | 167 | 40.46 | -25.54 |
| LIP/[E] | Karoo-Ferrar | 183.4 | 56.86 | -9.14 |
| LIP/E/I | CAMP/end-Triassic/Rochechouart | 202.1 | 75.56 | +9.56 |
| I | Manicouagan | 214.6 | 88.06 | +22.06 |
| I | Lake Saint Martin | 227.8 | 37.99 | -28.01 |
| LIP/E | Siberian Traps/end-Permian extinction | 251.9 | 62.09 | -3.91 |
| I | Araguainha | 254.7 | 64.89 | -1.11 |
| E | end-Capitanian | 260.3 | 70.49 | +4.49 |
| I | West Clearwater | 286.2 | 96.39 | +30.39 |
| E | end-Serpukhovian | 323.2 | 70.12 | +4.12 |
| E | Late Devonian | 358.9 | 42.55 | -23.45 |
| E | Frasnian-Famennian | 373.9 | 57.55 | -8.45 |
| I | Siljan | 380.9 | 64.55 | -1.45 |
| E | end-Ordovician | 445 | 65.38 | -0.62 |
| LIP/[E] | Kalkarindji | 510.7 | 67.81 | +1.81 |
| G | Gaskiers | 582.4 | 76.24 | +10.24 |
| G | Marinoan | 635.6 | 66.17 | +0.17 |
| LIP/G | Franklin | 715.4 | 82.7 | +16.7 |
| LIP | Guibei | 825 | 65.76 | -0.24 |
| LIP | Warakurna | 1076 | 63.68 | -2.32 |
| LIP | Umkondo | 1110 | 34.41 | -31.59 |
| LIP | Mackenzie | 1267 | 64.87 | -1.13 |
| I | Sudbury | 1849.3 | 77.74 | +11.74 |
| I | Vredefort | 2023 | 61.63 | -4.37 |

Table 1. Ages of events or event combinations in descending absolute age order. Event types are impact crater (I), large igneous province (LIP), most severe extinction (E) and Neoproterozoic glaciation (G). Two extinctions associated with LIPs but not amongst most severe are listed as [E].

Similarly, [GC] indicates global climate change within one million years of Ries impact. EA is the equivalent age based on modulo 63.27 Myr and ED is the difference from the assumed central value of 66.0 Ma. Here, EA is illustrated around 66.0 Ma, i.e., 66.0 – (63.27/2) to 66.0 + (63.27/2). Absolute age may be an average of two or three events or separate age estimates of the same event with overlapping errors. See Supplementary Information for details on event size, age errors and references.

| ED (Myr) | Most recent and predicted future | Predicted values in successive 63.27 Myr difference sequences (1 d.p.) | | | | | | | | | | Earlier ages closest to observed events | | Earliest solar system |
|---|---|---|---|---|---|---|---|---|---|---|---|---|---|---|
| 11.09 | *13.82* | 77.1 | 140.4 | 203.6 | 266.9 | 330.2 | 393.4 | 456.7 | 520.0 | 583.3 | 646.5 | | 1848.7 | 4569.3 |
| 10.2 | *12.93* | 76.2 | 139.5 | 202.7 | 266.0 | 329.3 | 392.6 | 455.8 | 519.1 | 582.4 | 645.6 | | | 4568.4 |
| 8.9 | *11.63* | 74.9 | 138.2 | 201.4 | 264.7 | 328.0 | 391.3 | 454.5 | 517.8 | 581.1 | 644.3 | | | **4567.1** |
| 4.516 | *7.246* | 70.5 | 133.8 | 197.1 | 260.3 | 323.6 | 386.9 | 450.1 | 513.4 | 576.7 | 639.9 | | | **4562.7** |
| 0 | 2.73 | *66.0* | 129.3 | 192.5 | 255.8 | 319.1 | 382.4 | 445.6 | 508.9 | 572.2 | 635.4 | 825.2 | 1268.1 | 4558.2 |
| -4.516 | -1.786 | 61.5 | 124.8 | 188.0 | 251.3 | 314.6 | 377.8 | 441.1 | 504.4 | 567.6 | 630.9 | | 2022.9 | |
| -8.9 | -6.17 | 57.1 | 120.4 | 183.6 | 246.9 | 310.2 | 373.5 | 436.7 | 500.0 | 563.3 | 626.5 | | | |
| -10.2 | -7.47 | 55.8 | 119.1 | 182.3 | 245.6 | 308.9 | 372.2 | 435.4 | 498.7 | 562.0 | 625.2 | | | |
| -11.09 | -8.36 | 54.9 | 118.2 | 181.5 | 244.7 | 308.0 | 371.3 | 434.5 | 497.8 | 561.1 | 624.3 | | | |

Table 2. Predicted event ages from a symmetrical model of 63.27 million year sequences. Italicized values are parameters, set according to (a) three consecutive stratigraphic boundaries (Messinian-Tortonian at 7.246 Ma, Tortonian-Serravallian at 11.63 Ma and the Serravallian-Langhian at 13.82 Ma), (b) an intermediate value in the Serravallian (12.93 Ma, 10.2 Myr from the centre) in agreement with two observed impacts and the Gaskiers glaciation and (c) the Cretaceous-Paleogene boundary (66.0 Ma), which locates the central sequence. The two earliest predicted values in bold are within 0.25 Myr of the observed earliest solar system ages (4562.49 and 4567.3 Ma, see text).

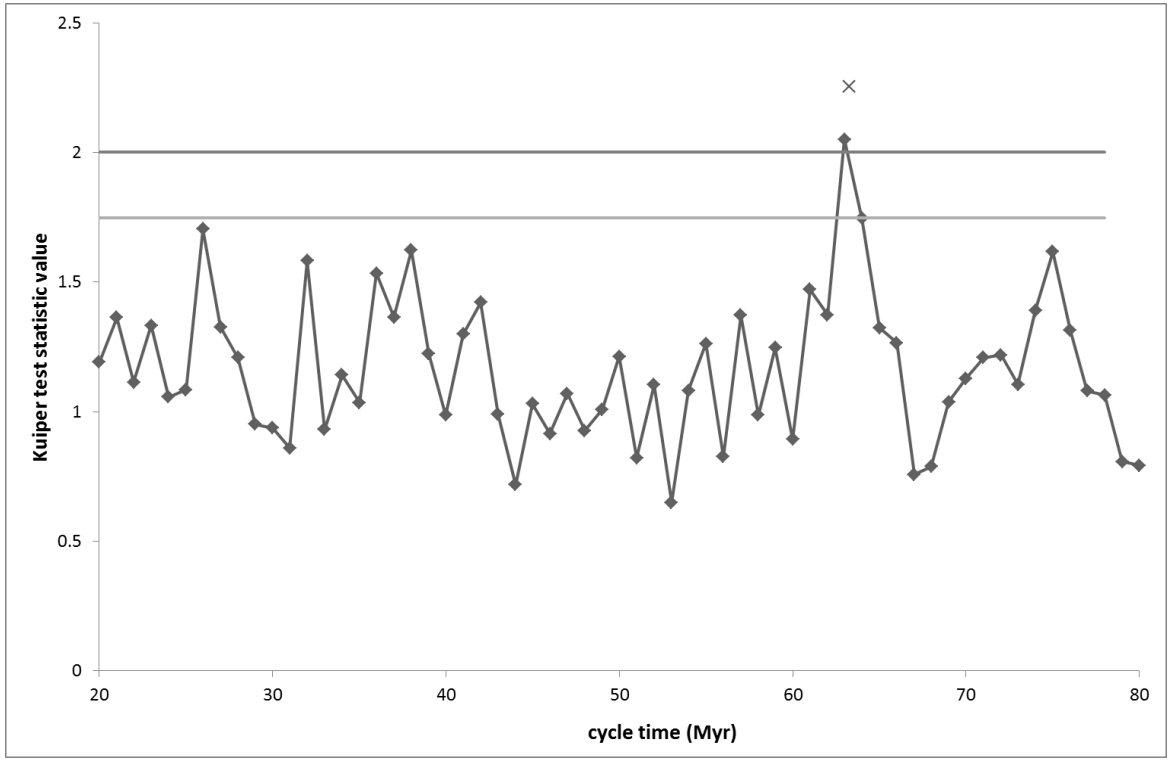

Figure 1. Change in Kuiper test statistic with cycle lengths from 20-80 Myr. The upper and lower horizontal lines are the P=0.01 and P=0.05 significance levels (*n*=38). The single point at the peak is 63.27 Myr with a test statistic value of 2.254 (measure of the maximum difference between null and observed distributions).

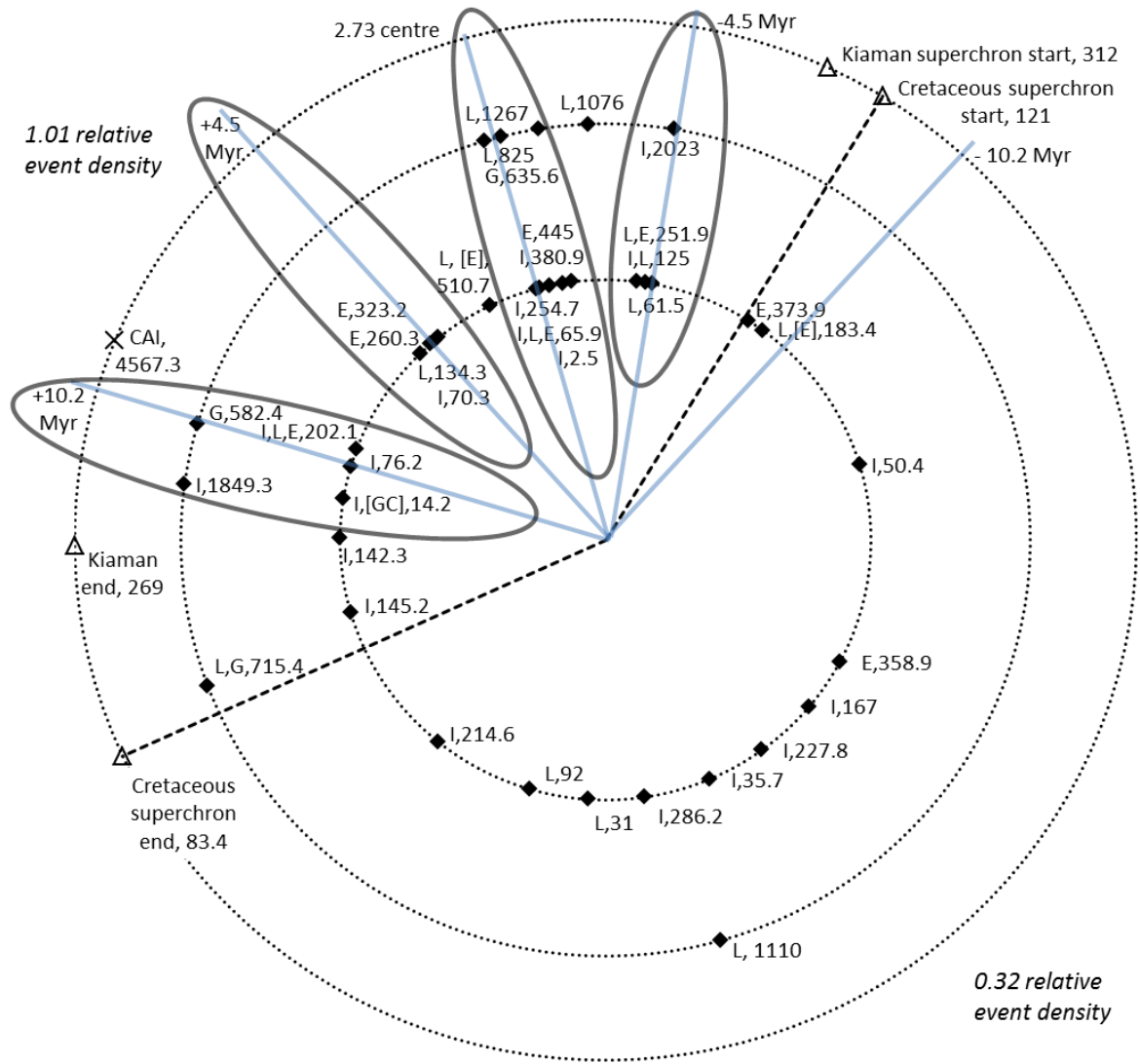

Figure 2. Distribution of globally disruptive events on a 63.27 Myr cycle. Significant clusters of events are indicated by ovals (see Table 2 in Supplementary Information). Events are shown as filled diamonds with event type abbreviated and followed by age (Ma, details in Table 1). The innermost dashed circle shows events in the Phanerozoic (541 Ma to present), the second dashed circle shows earlier events and the outer dashed circle gives information on superchron ages (open triangle), timings in the symmetrical model (Table 2) and the earliest Solar System age (CAI). The five solid radiating lines indicate the central values and two pairs of sequences differing by 4.5 and 10.2 Myr from the central value respectively. Relative event density is the number of events divided by the time within or outside of the equivalent ages of the Cretaceous superchron (shown as radiating dashed lines).